\documentclass[aps,prl,reprint,superscriptaddress,nofootinbib,longbibliography,floatfix,nobalancelastpage]{revtex4-2}
\usepackage{amsmath}
\usepackage{graphicx}
\usepackage[hidelinks]{hyperref}
\hypersetup{pdfauthor={Fu-Guo Yang},pdftitle={Tricritical Dynamics Reveal Black Hole Interior Geometry}}
\graphicspath{{figures/}}
\begin{document}
\raggedbottom
\title{Tricritical Dynamics Reveal Black Hole Interior Geometry}
\author{Fu-Guo Yang}
\email{yang-fuguo@outlook.com}
\affiliation{International Centre for Theoretical Physics Asia-Pacific, University of Chinese Academy of Sciences, Beijing 100190, China}
\begin{abstract}
We reconstruct the interior Kasner exponents of a scalarized black-hole equilibrium from the growth rate of a radial perturbation outside the horizon. The equal-entropy coexistence line of Reissner--Nordstr\"om and scalarized phases ends at a tricritical point. We follow the intermediate equilibrium separating the two phases along the coexistence line. The radial growth rate $\Omega_u$ is proportional to the magnitude of the entropy-potential curvature. As the tricritical point is approached, the curvature vanishes as the fourth power of the horizon scalar amplitude. The interior logarithmic scalar gradient grows inversely with the horizon amplitude. Eliminating the amplitude connects exterior dynamics to interior geometry, giving the angular Kasner exponent $p\sim2\sqrt{\Omega_u/C}$. The static source expansion and the critical Reissner--Nordstr\"om zero mode fix $C$ before the scalarized interior is integrated. We compute the growth rate from linear radial perturbations on asymptotically flat and anti-de Sitter black-hole backgrounds. At the sampled equilibria closest to tricriticality, the reconstructed angular exponent agrees with the result of independent inward integration at about the $0.5\%$ level.
\end{abstract}
\maketitle

Can the dynamics outside a black hole reveal its interior geometry near the singularity? In a Kasner regime, local spatial scale factors are powers of the proper time remaining to the singularity. The Kasner exponents describe the direction-dependent evolution of the geometry. We relate the Kasner exponents to the growth rate of an exterior radial scalar perturbation.

Charged black-hole scalarization provides a concrete setting. Charged black holes can develop scalar hair through a scalar instability of the Reissner--Nordstr\"om (RN) solution~\cite{Review,Herdeiro,Fernandes}. A hairy equilibrium with a growing radial mode can separate radially stable RN and hairy branches~\cite{HotCold}. RN and hairy branches also occur in asymptotically anti-de Sitter (AdS) spacetime~\cite{GuoAdSEMS,ChenAdSBranches}. Static source expansions describe competing scalarized phases~\cite{Muniz,FirstPrinciples}, and related holographic systems exhibit tricritical scaling~\cite{Matsumoto}.

Inside the horizon, scalar backreaction can replace the RN Cauchy horizon with a spacelike Kasner singularity~\cite{Hartnoll,Dias}. The terminal scalar gradient can grow inversely with the horizon amplitude~\cite{Interior,Scaling,Hierarchy}. Inverse-amplitude amplification also occurs in a unified description of stationary Kerr--Newman-like interiors with small scalar input~\cite{Unified}. A dynamical reconstruction also needs the dependence of the exterior growth rate on the horizon amplitude.

We approach the tricritical point along the equal-entropy coexistence line of RN and hairy equilibria. The coexisting phases have the same mass and charge. An intermediate equilibrium with a growing radial mode has lower entropy. Toward tricriticality, the intermediate equilibrium's growth rate decreases while the interior scalar gradient increases. Eliminating the horizon amplitude relates the growth rate to the Kasner exponents, with a coefficient fixed without scalarized-interior data. We test the reconstruction for asymptotically flat and AdS black holes. The End Matter gives the detailed static expansion, the interior coefficient, and the growth-rate normalization. The Supplemental Material gives numerical algorithms, independent coefficient checks, and error estimates~\cite{SM}.

\paragraph*{\normalfont\bfseries Model.} Einstein--Maxwell--scalar (EMS) gravity is described by the action
\begin{equation}
\begin{aligned}
I&=\frac1{16\pi}\int d^4x\sqrt{-\det(g_{ab})}\,\mathcal L,\\
\mathcal L&=R-2\Lambda-2g^{ab}\partial_a\psi\partial_b\psi-Z(\psi)F_{ab}F^{ab}.
\end{aligned}
\label{eq:lagrangian}
\end{equation}
Here $g_{ab}$ is the metric, $R$ the Ricci scalar, $\Lambda$ the cosmological constant, and $\psi$ a neutral massless scalar. We use $G=c=\hbar=k_B=1$. The electromagnetic potential $A_a$ defines $F_{ab}=\partial_aA_b-\partial_bA_a$. Flat asymptotics have $\Lambda=0$, and AdS asymptotics have $\Lambda=-3/L^2$, with curvature radius $L$. The coupling function $Z(\psi)$ is positive and even, with
\begin{equation}
\ln Z=a_2\psi^2+a_4\psi^4+a_6\psi^6.
\label{eq:coupling}
\end{equation}
The static spherical metric is
\begin{equation}
\begin{aligned}
ds^2&=-Ne^{-2\chi}dt^2+\frac{dr^2}{N}+r^2d\Omega_2^2,\\
N&=1-\frac{2m}{r}-\frac{\Lambda r^2}{3}.
\end{aligned}
\label{eq:metric}
\end{equation}
Here $r$ is the areal radius, $d\Omega_2^2$ the unit-sphere metric, and $m(r),\chi(r)$ the mass and redshift functions. A prime denotes $d/dr$. A comma followed by a variable denotes differentiation with respect to that variable. We fix $\chi(\infty)=0$, so in AdS the boundary metric is $-dt^2+L^2d\Omega_2^2$. The event horizon is the outer zero $r_H$ of $N$, the horizon scalar amplitude is $\varepsilon=\psi(r_H)>0$, and the mass is $M=m(\infty)$. For $A_a dx^a=A_t(r)dt$, Maxwell's equation gives the conserved electric charge:
\begin{equation}
(r^2e^\chi ZF_{rt})'=0,\quad Q=r^2e^\chi ZF_{rt}.
\label{eq:charge}
\end{equation}
After eliminating the electric field and $\chi'$ from the scalar equation, the remaining static equations are
\begin{equation}
\begin{gathered}
m'=\frac12\left[r^2N(\psi')^2+\frac{Q^2}{r^2Z}\right],\quad \chi'=-r(\psi')^2,\\
(r^2N\psi')'+r^3N(\psi')^3=\frac{Q^2}{2r^2}\partial_\psi Z^{-1}.
\end{gathered}
\label{eq:staticfield}
\end{equation}
For RN, $\psi=0$, and the electric contribution to $m(r)$ gives
\begin{equation}
m=M-\frac{Q^2}{2r},\quad N=1-\frac{2M}{r}+\frac{Q^2}{r^2}-\frac{\Lambda r^2}{3}.
\label{eq:rnlimit}
\end{equation}
The asymptotic scalar field defines the source $\psi_0$, scalar charge $Q_\psi$, and normalizable AdS coefficient $B$:
\begin{equation}
\begin{aligned}
\psi&=\psi_0+\frac{Q_\psi}{r}+\cdots\quad(\text{flat}),\\
\psi&=\psi_0+\frac{B}{r^3}+\cdots\quad(\text{AdS}).
\end{aligned}
\label{eq:tails}
\end{equation}
The equilibria of interest have $\psi_0=0$. The entropy and Hawking temperature are
\begin{equation}
S=\pi r_H^2,\quad T_H=\frac{e^{-\chi(r_H)}N'(r_H)}{4\pi}.
\label{eq:thermodynamicdata}
\end{equation}
At fixed $M,Q$, the first law gives $dS=\Pi\,d\psi_0$, with~\cite{Gibbons,deHaro,Astefanesei}
\begin{equation}
\Pi=\begin{cases}Q_\psi/T_H,&\text{flat},\\3B/(L^2T_H),&\text{AdS}.\end{cases}
\label{eq:conjugate}
\end{equation}
To compare the zero-source equilibria at fixed $M,Q$, we include nearby static solutions with small nonzero sources. We define the entropy potential
\begin{equation}
\Gamma=\psi_0\Pi-S,\quad d\Gamma=\psi_0d\Pi\quad(M,Q\text{ fixed}).
\label{eq:potential}
\end{equation}
At zero source, $\Gamma$ is stationary and equals $-S$. Minima with equal $\Gamma$ represent phases with equal entropy. The radial perturbation considered below preserves $M$ and $Q$ at linear order.

\paragraph*{\normalfont\bfseries Tricritical source coefficients.} Fix $M=M_c$, $\Lambda$, and $a_2$. The first RN charge admitting a positive nodeless static scalar mode with zero source defines the onset $Q_c$. The critical RN horizon radius is $r_c$. To calculate the source coefficients, first expand at fixed $r_H=r_c$, $Q=Q_c$:
\begin{equation}
\psi=\varepsilon\phi_1+\varepsilon^3\phi_3+\varepsilon^5\phi_5+\cdots.
\label{eq:modeexpansion}
\end{equation}
The leading mode has $\phi_1(r_c)=1$. The metric responses are derived in End Matter, Sec.~\ref{em:tricritical}. At first order, the scalar equation gives
\begin{equation}
\begin{gathered}
N_0=1-\frac{2M_c}{r}+\frac{Q_c^2}{r^2}-\frac{\Lambda r^2}{3},\\
\mathcal L_0\phi_1\equiv(r^2N_0\phi_1')'+\frac{a_2 Q_c^2}{r^2}\phi_1=0.
\end{gathered}
\label{eq:criticalmode}
\end{equation}
We impose $\phi_1(\infty)=0$ and choose a nonextremal onset. Among the coupling coefficients, only $a_2$ enters Eq.~\eqref{eq:criticalmode}. The critical RN temperature is $T_c=N_0'(r_c)/(4\pi)$. At large $r$, the critical mode has the form
\begin{equation}
\phi_1\sim\frac{\sigma r_c}{r}\quad(\text{flat}),\quad \phi_1\sim\frac{A}{r^3}\quad(\text{AdS}).
\label{eq:criticaltails}
\end{equation}
The scalar tails give $\Pi=\kappa\varepsilon+\cdots$, where
\begin{equation}
\kappa=\frac1{T_c}\begin{cases}\sigma r_c,&\text{flat},\\3A/L^2,&\text{AdS},\end{cases}\quad \kappa>0.
\label{eq:kappa}
\end{equation}

For the linear charge response, solve the horizon-regular mode $\phi_1(r;Q)$ on RN backgrounds at fixed $M=M_c$, with $\phi_1(r_H(Q);Q)=1$. The source derivative gives
\begin{equation}
\alpha=Q_c\left.\partial_Q\phi_1(\infty;Q)\right|_{Q_c}<0.
\label{eq:linearsource}
\end{equation}
Let $g$ be the cubic source coefficient and $a_*$ the coupling where $g=0$. The quintic coefficient at $a_4=a_*$ is $v_*$. Retaining the leading tricritical orders, the boundary source has the form
\begin{equation}
\psi_0=\alpha\Delta\varepsilon+g\varepsilon^3+v_*\varepsilon^5+\cdots,\quad \Delta=Q/Q_c-1.
\label{eq:source}
\end{equation}
Zero-mode projection gives
\begin{equation}
a_*=\frac{a_2^2}{2}-\frac{\displaystyle\int_{r_c}^{\infty}r^3N_0\phi_1(\phi_1')^3\,dr}{\displaystyle Q_c^2\int_{r_c}^{\infty}\phi_1^4r^{-2}\,dr},
\label{eq:tricriticalcoupling}
\end{equation}
The cubic coefficient is
\begin{equation}
g=-\frac{2Q_c^2(a_4-a_*)}{\kappa T_c}\int_{r_c}^{\infty}\frac{\phi_1^4}{r^2}\,dr.
\label{eq:cubiccoefficient}
\end{equation}
The integral is positive, so increasing $a_4$ through $a_*$ changes $g$ from positive to negative. We keep $a_2,a_6$ fixed during this tuning. The fifth-order calculation gives $v_*>0$ at $a_4=a_*$ for the families studied here. End Matter, Sec.~\ref{em:tricritical}, gives the detailed coefficient calculation.

Before using the source coefficients in the entropy potential at fixed $M,Q$, we convert the fixed-area calculation to fixed mass. The first law at fixed $Q$ is
\begin{equation}
dM=T_HdS-T_H\Pi\,d\psi_0.
\label{eq:firstlaw}
\end{equation}
At $Q=Q_c$ and fixed horizon area, insert $\Pi=\kappa\varepsilon+\cdots$ and $\psi_0=\phi_3(\infty)\varepsilon^3+\cdots$, then integrate in $\varepsilon$:
\begin{equation}
M-M_c=-\frac34\kappa T_c\phi_3(\infty)\varepsilon^4+O(\varepsilon^6).
\label{eq:massshift}
\end{equation}
On critical RN, $\partial_{r_H}M=2\pi r_cT_c>0$. Holding $M=M_c$ therefore requires changing $r_H$. Since $\psi_{0,r_H}=O(\varepsilon)$, the required shifts are
\begin{equation}
\begin{aligned}
\delta r_H&=O(|\phi_3(\infty)|\varepsilon^4+\varepsilon^6),\\
\delta\psi_0&=O(|\phi_3(\infty)|\varepsilon^5+\varepsilon^7).
\end{aligned}
\label{eq:sourceshift}
\end{equation}
The cubic coefficient is unchanged, as is the quintic coefficient when $\phi_3(\infty)=0$. Thus Eq.~\eqref{eq:source} applies at fixed $M,Q$ with $g=\phi_3(\infty)$ and $v_*=\phi_5(\infty)|_{a_4=a_*}$. The coupling $a_6$ first enters at fifth order and changes $v_*$ while leaving $Q_c,a_*$ unchanged.

\paragraph*{\normalfont\bfseries Entropy coexistence and slowing.} Since $\Pi_{,\varepsilon}\to\kappa>0$, the horizon amplitude $\varepsilon$ parametrizes the entropy potential near onset. Keeping the leading tricritical orders, integration of Eq.~\eqref{eq:potential} gives
\begin{equation}
\frac{\Gamma-\Gamma_{\rm RN}}{\kappa}
=\frac{\alpha\Delta}{2}\varepsilon^2+\frac g4\varepsilon^4+\frac{v_*}{6}\varepsilon^6+\cdots.
\label{eq:entropyexpansion}
\end{equation}
Here $\Gamma_{\rm RN}=-S_{\rm RN}$ at the same $M,Q$. For $g>0$, a nonzero minimum emerges continuously as $\Delta$ increases through zero, with $\varepsilon^2\sim-\alpha\Delta/g$. At $g=0$, the onset instead has $\varepsilon^4\sim-\alpha\Delta/v_*$. The simultaneous vanishing of the quadratic and quartic terms, with a positive sixth-order term, defines the tricritical point $(Q_c,a_*)$.

For $g<0$, equal-entropy RN and scalarized equilibria occur at $\Delta_{\rm eq}<0$, before the linear RN onset. We label the larger-amplitude scalarized coexistence equilibrium by eq and the intermediate equilibrium by $u$. Equal entropy in Eq.~\eqref{eq:entropyexpansion} and the zero-source condition in Eq.~\eqref{eq:source} require
\begin{equation}
\begin{aligned}
\alpha\Delta_{\rm eq}+g\varepsilon_{\rm eq}^2+v_*\varepsilon_{\rm eq}^4&=0,\\
\frac{\alpha\Delta_{\rm eq}}2+\frac g4\varepsilon_{\rm eq}^2+\frac{v_*}{6}\varepsilon_{\rm eq}^4&=0.
\end{aligned}
\label{eq:coexistenceconditions}
\end{equation}
Eliminating $\Delta_{\rm eq}$ gives the larger coexistence amplitude. Dividing the zero-source equation by $\varepsilon$ gives a quadratic in $\varepsilon^2$, whose roots sum to $-g/v_*$. The smaller root is the intermediate equilibrium. Thus
\begin{equation}
\varepsilon_{\rm eq}^2\sim-\frac{3g}{4v_*},\quad
\varepsilon_u^2\sim-\frac{g}{4v_*},\quad
\alpha\Delta_{\rm eq}\sim\frac{3g^2}{16v_*}.
\label{eq:coex}
\end{equation}
The intermediate equilibrium is a maximum of $\Gamma$ and has lower entropy than either minimum. RN, $u$, and the larger-amplitude scalarized equilibrium have the same mass, charge, and couplings. We follow $u$ along coexistence as $a_4\to a_*^+$, tuning the charge at each step. In Eq.~\eqref{eq:coex} and the subsequent coexistence scaling relations, $\sim$ denotes a ratio tending to one in this limit. Along coexistence, $g=O(\varepsilon_u^2)$ and $\Delta_{\rm eq}=O(\varepsilon_u^4)$, so corrections to $\Pi_{,\varepsilon}=\kappa$ affect the potential only beyond sixth order.

At the intermediate equilibrium, differentiate with respect to $\varepsilon$ at fixed $M,Q$ and couplings. Since $\psi_0=0$, Eq.~\eqref{eq:potential} gives the normalized entropy curvature
\begin{equation}
K_u\equiv\left.\frac{\Gamma_{,\varepsilon\varepsilon}}{\kappa}\right|_u
=\left.\frac{\Pi_{,\varepsilon}}{\kappa}\psi_{0,\varepsilon}\right|_u
\sim\left.\psi_{0,\varepsilon}\right|_u.
\label{eq:curvaturesource}
\end{equation}
The zero-source condition gives $\alpha\Delta_{\rm eq}=-g\varepsilon_u^2-v_*\varepsilon_u^4$. Substituting into the source slope and using Eq.~\eqref{eq:coex} gives
\begin{equation}
\begin{aligned}
K_u&\sim\alpha\Delta_{\rm eq}+3g\varepsilon_u^2+5v_*\varepsilon_u^4\\
&\sim2g\varepsilon_u^2+4v_*\varepsilon_u^4\sim-4v_*\varepsilon_u^4<0.
\end{aligned}
\label{eq:potentialcurvature}
\end{equation}
The fourth-power dependence follows because $g\sim-4v_*\varepsilon_u^2$ also tends to zero as the tricritical endpoint is approached along coexistence.

A radial perturbation has time dependence $\delta\psi=e^{\Omega t}\Psi(r)/r$, where $\Psi$ is the radial master field and $\Omega>0$ the growth rate. Define the tortoise coordinate $r_*$ by $dr_*/dr=e^\chi/N$. Eliminating the metric and electric perturbations gives~\cite{HotCold,GuoAdSEMS}
\begin{equation}
[-\partial_{r_*}^2+V]\Psi=-\Omega^2\Psi.
\label{eq:radialmaster}
\end{equation}
The real potential $V$ vanishes at a regular nonextremal horizon. The ingoing condition gives $\Psi\propto e^{\Omega r_*}$ as $r_*\to-\infty$. At flat infinity, $V\to0$ and the outgoing condition gives $\Psi\propto e^{-\Omega r_*}$. The zero-source AdS condition gives $\Psi\propto r^{-2}$. These boundary conditions select the growing rate $\Omega_u$ on the intermediate equilibrium.
At zero growth rate, $rw(r;0)$ solves Eq.~\eqref{eq:radialmaster}, where
\begin{equation}
\begin{aligned}
w(r;0)&=\left.\partial_\varepsilon\psi\right|_{r_H,Q},\\
w(r_H;0)&=1,\quad w_0=w(\infty;0).
\end{aligned}
\label{eq:jacobi}
\end{equation}
Its boundary source $w_0$ need not vanish. The fixed-mass differentiation in End Matter gives $w_0\sim\psi_{0,\varepsilon}|_{M,Q}\sim K_u$.

Near onset, the radial calculation relates the source slope to the growth rate through a positive coefficient $\mathcal D$:
\begin{gather}
\Omega_u\sim-\mathcal D\left.\psi_{0,\varepsilon}\right|_u
\sim-\mathcal D K_u\sim4\mathcal Dv_*\varepsilon_u^4,
\label{eq:growth}\\
\mathcal D_{\rm flat}=\frac{\sigma}{r_c(1+\sigma^2)},\quad
\mathcal D_{\rm AdS}=\frac{3A}{L^2r_c^2}.
\label{eq:thresholdcoefficients}
\end{gather}
Tricritical slowing denotes the vanishing radial growth rate as the tricritical point is approached along coexistence:
\begin{equation}
\Omega_u\propto(a_4-a_*)^2\longrightarrow0^+,\quad a_4\to a_*^+.
\label{eq:tricriticalslowing}
\end{equation}
The growth-rate relation and the coefficient $\mathcal D$ are derived in End Matter, Sec.~\ref{em:growth}.

\paragraph*{\normalfont\bfseries Dynamical reconstruction of the Kasner exponents.} Let $\beta_u=-\lim_{r\to0}r\partial_r\psi_u$ be the logarithmic scalar gradient near the singularity. For the equilibria studied here, small-hair matching gives the leading inverse-amplitude law~\cite{Scaling,Hierarchy,Unified,HoloCauchy}
\begin{equation}
\beta_u\sim\frac1{b\varepsilon_u}.
\label{eq:amplification}
\end{equation}
The coefficient $b$ is fixed by the first-order critical RN mode. End Matter, Sec.~\ref{em:tricritical}, gives $b$ for both asymptotics.
The scalar--Kasner relations give the angular and longitudinal exponents:
\begin{equation}
p=\frac2{\beta_u^2+3},\quad p_\parallel=1-2p.
\label{eq:beta}
\end{equation}

Let $\tau$ be the proper time remaining to the singularity. Inside the horizon, $x_\parallel$ is proportional to $t$, and $x_1,x_2$ describe the local angular directions. Up to constant spatial rescalings, the local metric takes the scalar--Kasner form
\begin{equation}
ds^2\simeq-d\tau^2+\tau^{2p_\parallel}dx_\parallel^2
+\tau^{2p}(dx_1^2+dx_2^2).
\label{eq:kasnermetric}
\end{equation}
We extract $\beta_u$ by inward integration as $r\to0$ at each fixed $\varepsilon_u>0$, then take $\varepsilon_u\to0$ along coexistence.

Eliminating the horizon amplitude between Eqs.~\eqref{eq:growth} and \eqref{eq:amplification} gives the prediction along coexistence as $g\to0^-$
\begin{equation}
\boxed{\Omega_u\beta_u^4\to C,\quad C=\frac{4\mathcal Dv_*}{b^4}.}
\label{eq:timegeometry}
\end{equation}
The coefficient $C$ has units of inverse time. The zero-mode expansion fixes $v_*$, and the critical mode fixes $\mathcal D$ and $b$. Varying $a_6$ at fixed $a_2$ leaves the critical RN mode unchanged and changes $C$ through $v_*$.

Inserting $\beta_u\sim(C/\Omega_u)^{1/4}$ into Eq.~\eqref{eq:beta} gives the angular estimate $p_{\rm dyn}$. Define the e-folding time by $t_u\equiv\Omega_u^{-1}$.
\begin{equation}
\boxed{p_{\rm dyn}=\frac{2}{3+\sqrt{C/\Omega_u}}=\frac{2}{3+\sqrt{Ct_u}}.}
\label{eq:geometryreadout}
\end{equation}
The longitudinal estimate is $1-2p_{\rm dyn}$. Toward tricriticality, slower exterior growth corresponds to $p\to0$ and $p_\parallel\to1$. The normalizations of the spatial scale factors require additional interior data.

\begin{figure}[hbp]
\centering\includegraphics[width=\columnwidth]{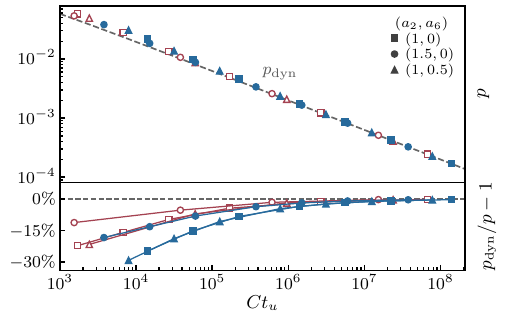}
\caption{Dynamical reconstruction for flat (open red) and AdS (filled blue) equilibria. Marker shapes identify $(a_2,a_6)$. Symbols show the directly integrated exponent $p$ against $Ct_u=C/\Omega_u$, with $\Omega_u$ computed from exterior linear perturbations. The dashed curve is Eq.~\eqref{eq:geometryreadout}. The lower panel shows the relative difference $(p_{\rm dyn}-p)/p$, expressed as a percentage.}
\label{fig:dynamic}
\end{figure}

\paragraph*{\normalfont\bfseries Discussion and conclusions.} Figure~\ref{fig:dynamic} tests the reconstruction against independent inward integrations for three coupling families with each choice of asymptotics. For AdS, $a_*,v_*$ are obtained numerically from exterior source fits and independently checked against the zero-mode expansion. At the sampled equilibria closest to tricriticality, the relative reconstruction discrepancy is below about $0.5\%$ and decreases toward the tricritical point.

A possible AdS measurement would apply a weak boundary source and read the scalar response after switching the source off~\cite{SonStarinets}. For the intermediate equilibrium with static coefficient $B_u$, let $\delta B(t)=B(t)-B_u$. During a linear interval dominated by the growing mode,
\begin{equation}
\delta B(t)\propto e^{\Omega_u t},\quad \Omega_u=\partial_t\ln|\delta B|.
\label{eq:boundarymeasurement}
\end{equation}
For flat asymptotics, the analogous signal would be outgoing scalar radiation. An astrophysical measurement requires a black hole close to the intermediate equilibrium, excitation of the growing mode, and a detectable scalar signal~\cite{GWReview,EMSQNM,GWScalarization,ScalarRingdown}. Closer to tricriticality the reconstruction is more accurate, but the slower mode amplifies less over a fixed observing interval.

High-overtone probe spectra offer a possible independent route to the interior exponents~\cite{HearKasner,DecodeKasner}. For our spherical AdS equilibria, one would compute the high-overtone quasinormal frequencies of a neutral, massless, minimally coupled scalar at several angular harmonic numbers. If a spectral correction proportional to $n^{-\delta_u}$, with overtone number $n$, yields $p_{\rm spec}=1/(1+\delta_u)$, combining the spectral and dynamical reconstructions gives the exterior constraint
\begin{equation}
\Omega_u(2\delta_u-1)^2\longrightarrow C.
\label{eq:externalconstraint}
\end{equation}
Identifying and extracting the $n^{-\delta_u}$ spectral correction in the EMS families studied here is a future task. Other critical branches may also permit reconstruction, once the growth or decay rate and terminal scalar gradient are known as functions of the same branch parameter~\cite{HohenbergHalperin}.

\paragraph*{\normalfont\bfseries Acknowledgments.} The author gratefully acknowledges the assistance of OpenAI's ChatGPT with numerical programming, checking the derivations, and polishing the manuscript. The author takes full responsibility for all scientific results and the content of this paper.

\clearpage
\section*{End Matter}
\appendix
\setcounter{secnumdepth}{1}
\section{Static expansion about critical RN}\label{em:tricritical}
For the scalar expansion in Eq.~\eqref{eq:modeexpansion}, we first hold $r_H=r_c$, $Q=Q_c$. The conversion to fixed mass is given in the main text. The metric responses are
\begin{equation}
\begin{aligned}
m&=M_c-\frac{Q_c^2}{2r}+\varepsilon^2m_2+\varepsilon^4m_4+O(\varepsilon^6),\\
\chi&=\varepsilon^2\chi_2+\varepsilon^4\chi_4+O(\varepsilon^6).
\end{aligned}
\label{em:expansion}
\end{equation}
The regular corrections $\phi_3,\phi_5,m_2,m_4$ vanish at $r_c$. The redshift corrections follow from Eq.~\eqref{eq:staticfield} with $\chi(\infty)=0$.

The coupling term in the scalar equation expands as
\begin{equation}
\begin{aligned}
\frac12\partial_\psi Z^{-1}={}&-a_2\psi+(a_2^2-2a_4)\psi^3\\
&+\left(3a_2a_4-\frac{a_2^3}{2}-3a_6\right)\psi^5+O(\psi^7).
\end{aligned}
\label{em:couplingresponse}
\end{equation}

Collecting equal powers of $\varepsilon$ in Eq.~\eqref{eq:staticfield} gives
\begin{equation}
\mathcal L_0\phi_k=S_k,\quad k=1,3,5,
\label{em:hierarchy}
\end{equation}
with the critical RN operator $\mathcal L_0$ in Eq.~\eqref{eq:criticalmode}. Each higher-order source $S_k$ depends on previously determined fields. For $k=3,5$, subtracting $\phi_k\mathcal L_0\phi_1=0$ from $\phi_1\mathcal L_0\phi_k=\phi_1S_k$ gives
\begin{equation}
\begin{aligned}
{}[r^2N_0(\phi_1\phi_k'-\phi_k\phi_1')]'&=\phi_1S_k,\\
\phi_k(\infty)&=\frac{\int_{r_c}^{\infty}\phi_1S_k\,dr}{\kappa T_c}.
\end{aligned}
\label{em:projection}
\end{equation}
The horizon term vanishes. At infinity, both asymptotics give $\kappa T_c\phi_k(\infty)$ by Eq.~\eqref{eq:kappa}. 
\paragraph*{\normalfont\bfseries First order.} Near the critical RN inner horizon $r_I$, the linear scalar mode has logarithmic behavior,
\[
\phi_1(r)=-b\ln\!\frac{r-r_I}{r_I}+O(1),\quad r\to r_I^+.
\]
Integrating Eq.~\eqref{eq:criticalmode} between $r_c$ and $r_I$ gives the coefficient in Eq.~\eqref{eq:amplification}:
\begin{equation}
b=-\frac{a_2 Q_c^2}{r_I^2N_0'(r_I)}\int_{r_I}^{r_c}\frac{\phi_1(r)}{r^2}\,dr.
\label{eq:innercoefficient}
\end{equation}

\paragraph*{\normalfont\bfseries Third order.} Using $\mathcal L_0\phi_1=0$, the second-order constraint integrates directly:
\begin{equation}
m_2=\frac12r^2N_0\phi_1\phi_1',\quad m_2(r_c)=m_2(\infty)=0.
\label{em:secondorder}
\end{equation}
Thus the mass has no quadratic shift at fixed horizon area. The cubic equation is $\mathcal L_0\phi_3=S_3$, with
\begin{equation}
S_3=\phi_1[r^3N_0(\phi_1')^2]'+\frac{Q_c^2(a_2^2-2a_4)}{r^2}\phi_1^3.
\label{em:cubicsource}
\end{equation}
The cubic boundary source vanishes when $\int_{r_c}^{\infty}\phi_1S_3\,dr=0$. Integrating by parts gives the coupling and cubic response in Eqs.~\eqref{eq:tricriticalcoupling} and \eqref{eq:cubiccoefficient}.
The integration boundary term vanishes for both asymptotics.

\paragraph*{\normalfont\bfseries Fifth order.} The fourth-order constraint, simplified with $\mathcal L_0\phi_1=0$, gives
\begin{equation}
\begin{aligned}
m_4'={}&(r^2N_0\phi_1'\phi_3)'-\frac12r^3N_0\phi_1(\phi_1')^3\\
&+\frac{Q_c^2(a_2^2-2a_4)}{4r^2}\phi_1^4.
\end{aligned}
\label{em:fourthorder}
\end{equation}
Integrate with $m_4(r_c)=0$. The quintic equation is $\mathcal L_0\phi_5=S_5$, where
\begin{equation}
\begin{aligned}
S_5={}&[2r(m_2\phi_3'+m_4\phi_1')]'\\
&-3r^3N_0(\phi_1')^2\phi_3'+2r^2m_2(\phi_1')^3\\
&+\frac{3Q_c^2(a_2^2-2a_4)}{r^2}\phi_1^2\phi_3\\
&+\frac{Q_c^2}{r^2}\left(3a_2 a_4-\frac{a_2^3}{2}-3a_6\right)\phi_1^5.
\end{aligned}
\label{em:quinticsource}
\end{equation}
At $a_4=a_*$, Eq.~\eqref{em:projection} with $k=5$ gives the quintic boundary source. Only $\phi_1,\phi_3,m_2,m_4$ enter $S_5$, so $\phi_5$ need not be solved separately.

\section{From entropy curvature to the radial growth rate}\label{em:growth}
A static variation at fixed $r_H,Q$ obeys the same metric constraint as the radial mode, $\delta m=r^2N\psi'\delta\psi$. Equation~\eqref{eq:firstlaw} gives
\begin{equation}
\begin{aligned}
M_{,\varepsilon}&=-T_H\Pi w_0,\\
M_{,r_H}&=2\pi r_HT_H-T_H\Pi\psi_{0,r_H}.
\end{aligned}
\label{em:massderivatives}
\end{equation}
The first derivative holds $r_H,Q$ fixed, and the second holds $\varepsilon,Q$ fixed. Adjusting $r_H$ to hold $M$ fixed gives
\begin{equation}
\begin{aligned}
\left.\partial_\varepsilon\psi_0\right|_{M,Q}
&=w_0-\frac{\psi_{0,r_H}M_{,\varepsilon}}{M_{,r_H}}\\
&=\frac{2\pi r_HT_H}{M_{,r_H}}w_0=[1+O(\varepsilon^2)]w_0.
\end{aligned}
\label{em:ensembleconversion}
\end{equation}
The correction starts at $\varepsilon^2$ because $\Pi$ and $\psi_{0,r_H}$ are both $O(\varepsilon)$. Equation~\eqref{eq:curvaturesource} then gives $w_0\sim K_u$.

For flat asymptotics, let $\Psi_-(r;\Omega)$ and $\Psi_+(r;\Omega)$ solve Eq.~\eqref{eq:radialmaster} and decay at the horizon and infinity, respectively. Normalize both solutions to one at a fixed exterior radius $r_m$, with tortoise coordinate $r_{*m}$. A mode satisfies both boundary conditions when the derivatives also agree:
\begin{equation}
\mathcal M(\Omega)=\partial_{r_*}\Psi_-(r_m;\Omega)-\partial_{r_*}\Psi_+(r_m;\Omega)=0.
\label{em:mismatch}
\end{equation}
Differentiate Eq.~\eqref{eq:radialmaster} with respect to $\Omega$ at fixed background. Multiplication by $\Psi_\pm$ and subtraction of the original equation multiplied by $\partial_\Omega\Psi_\pm$ gives
\begin{equation}
\begin{aligned}
W_\pm&=\Psi_\pm\partial_{r_*}\partial_\Omega\Psi_\pm-(\partial_{r_*}\Psi_\pm)\partial_\Omega\Psi_\pm,\\
\partial_{r_*}W_\pm&=2\Omega\Psi_\pm^2.
\end{aligned}
\label{em:wronskianvariation}
\end{equation}
The fixed normalization gives $\partial_\Omega\Psi_\pm(r_m;\Omega)=0$. Exponential decay at $\Omega>0$ gives $W_-(-\infty)=W_+(\infty)=0$. Integrating from each remote boundary to $r_{*m}$ and subtracting therefore yields
\begin{equation}
\partial_\Omega\mathcal M=2\Omega\left[\int_{-\infty}^{r_{*m}}\Psi_-^2dr_*+\int_{r_{*m}}^\infty\Psi_+^2dr_*\right].
\label{em:matching}
\end{equation}
At critical RN, both zero-rate solutions equal $\Psi_1(r)/\Psi_1(r_m)$, where $\Psi_1=r\phi_1$. The unnormalized field approaches $r_c$ at the horizon and $r_c\sigma$ at infinity. Each exponential tail contributes its limiting amplitude squared divided by $2\Omega$. Integrating at $\Omega>0$ before taking the limit gives
\begin{equation}
\partial_\Omega\mathcal M\longrightarrow
\frac{r_c^2}{\Psi_1(r_m)^2}+\frac{r_c^2\sigma^2}{\Psi_1(r_m)^2}
=\frac{r_c^2(1+\sigma^2)}{\Psi_1(r_m)^2}.
\label{em:matchinglimits}
\end{equation}
Any fixed finite interval contributes zero in this limit. The two $1/\Omega$ tails supply $1+\sigma^2$, making the mismatch linear in $\Omega$ despite the $\Omega^2$ term in the radial equation~\cite{KlausSimon,Golovaty,Egorova2015}.

At zero growth rate, the finite limit $\Psi_+(\infty;0)$ gives a zero-source scalar perturbation $\Psi_+(r;0)/r$. The horizon-regular master field is $rw(r;0)$. At infinity,
\[
\begin{gathered}
rw(r;0)=w_0r+O(1),\quad \partial_{r_*}[rw(r;0)]\to w_0,\\
\Psi_+(r;0)\to\Psi_+(\infty;0),\quad \partial_{r_*}\Psi_+(r;0)=O(r^{-2}).
\end{gathered}
\]
The constant Wronskian of the two master fields is therefore $\Psi_+(\infty;0)w_0$. Evaluation at $r_m$, with the fixed normalization, gives
\begin{equation}
\mathcal M(0)=\frac{\Psi_+(\infty;0)w_0}{r_mw(r_m;0)}.
\label{em:staticmismatch}
\end{equation}
Using the critical-mode normalization and tails in Eq.~\eqref{em:staticmismatch} gives $\mathcal M(0)/w_0\to r_c\sigma/\Psi_1(r_m)^2$. Along the nearby equilibrium branch, the growth-rate response approaches Eq.~\eqref{em:matchinglimits}. Integrating $\partial_\Omega\mathcal M$ from zero to $\Omega>0$ then gives
\begin{multline}
\frac{\Psi_1(r_m)^2}{r_c\sigma}\mathcal M(\Omega)=w_0\\
+\frac{r_c(1+\sigma^2)}{\sigma}\Omega+o(|w_0|+r_c\Omega).
\label{em:flatthreshold}
\end{multline}
Setting $\mathcal M(\Omega_u)=0$ gives $\Omega_u\sim-\mathcal D_{\rm flat}w_0$, with $\mathcal D_{\rm flat}$ in Eq.~\eqref{eq:thresholdcoefficients}.

For AdS asymptotics, write $w(r;\Omega)=\Psi(r)/r$ on critical RN. Equation~\eqref{eq:radialmaster} becomes
\begin{equation}
\mathcal L_0w=\Omega^2\frac{r^2}{N_0}w,
\label{em:adsscalarequation}
\end{equation}
where $\mathcal L_0$ is defined in Eq.~\eqref{eq:criticalmode}. Normalize $we^{-\Omega r_*}\to1$ at the horizon, so $w(r;0)=\phi_1(r)$. At $\Omega=0$, differentiating Eq.~\eqref{em:adsscalarequation} gives $\mathcal L_0w_{,\Omega}=0$. The two homogeneous equations give the constant weighted Wronskian:
\begin{equation}
\partial_r\!\left\{r^2N_0\left[\phi_1\partial_r w_{,\Omega}-\phi_1'w_{,\Omega}\right]\right\}=0.
\label{em:adswronskianidentity}
\end{equation}
On critical RN, $\chi=0$. Near the nonextremal horizon, $N_0=N_0'(r_c)(r-r_c)+O((r-r_c)^2)$. Integrating the tortoise-coordinate definition gives
\[
\begin{gathered}
\frac{dr_*}{dr}=\frac1{N_0}=\frac1{N_0'(r_c)(r-r_c)}+O(1),\\
r_*=\frac{\ln[(r-r_c)/r_c]}{N_0'(r_c)}+O(1).
\end{gathered}
\]
The ingoing solution is $w=e^{\Omega r_*}[1+O(r-r_c)]$. Its regular Frobenius factor has coefficients smooth in $\Omega$, and $w(r;0)=\phi_1(r)$. Differentiate at fixed $r>r_c$, set $\Omega=0$, and then take $r\to r_c^+$:
\begin{equation}
\begin{aligned}
\left.\partial_\Omega w\right|_0
&=r_*\phi_1+O(r-r_c)=r_*+O(1),\\
\left.\partial_r\partial_\Omega w\right|_0
&=\frac{\phi_1}{N_0}+r_*\phi_1'+O(1)\sim\frac1{N_0}.
\end{aligned}
\label{em:adshorizonderivatives}
\end{equation}
Here $\phi_1(r_c)=1$ and $\phi_1'$ is finite. The pole $1/N_0$ dominates the logarithmic term $r_*\phi_1'$. In Eq.~\eqref{em:adswronskianidentity}, the first Wronskian term tends to $r_c^2$; the second vanishes because $N_0r_*\to0$.

At the AdS boundary, the source is $w(\infty;\Omega)$, with $w(\infty;0)=0$. The zero-rate expansions are
\begin{equation}
\begin{gathered}
\phi_1\sim\frac{A}{r^3},\quad \phi_1'\sim-\frac{3A}{r^4},\\
\left.\partial_\Omega w\right|_0=w_{,\Omega}(\infty;0)+O(r^{-3}).
\end{gathered}
\label{em:adsboundaryexpansion}
\end{equation}
Since $r^2N_0\sim r^4/L^2$, the first Wronskian term vanishes and the negative of the second tends to $3Aw_{,\Omega}(\infty;0)/L^2$. Equating the horizon and boundary values gives
\begin{equation}
w_{,\Omega}(\infty;0)=\frac{L^2r_c^2}{3A}=\frac1{\mathcal D_{\rm AdS}}.
\label{em:adsthreshold}
\end{equation}
For nearby scalarized equilibria, $w_{,\Omega}(\infty;0)\to1/\mathcal D_{\rm AdS}$. The zero-source condition is
\begin{equation}
0=w_0+\frac{\Omega_u}{\mathcal D_{\rm AdS}}+o\!\left(|w_0|+\frac{\Omega_u}{\mathcal D_{\rm AdS}}\right),
\label{em:adssourcecondition}
\end{equation}
which gives $\Omega_u\sim-\mathcal D_{\rm AdS}w_0$.

Combining the two threshold results with Eqs.~\eqref{em:ensembleconversion} and \eqref{eq:potentialcurvature} gives the growth-rate relation in Eq.~\eqref{eq:growth}.

\end{document}